\documentclass[twocolumn]{aa}
\pdfoutput=1
\usepackage{natbib}
\bibpunct{(}{)}{;}{a}{}{,} 

\usepackage{graphicx}
\usepackage{txfonts}
\usepackage{upgreek}
\usepackage{comment}

\usepackage{tabularx} 
\usepackage{caption}
\usepackage{subfig}

\newcommand{\days}{\textrm{ days}}

\newcommand{\Kelvin}{\textrm{ K}}

\newcommand{\nm}{\textrm{ nm}}
\newcommand{\um}{\textrm{ } \upmu \textrm{m}}

\newcommand{\km}{\textrm{ km}}

\newcommand{\pc}{\textrm{ pc}}

\newcommand{\HTWO}{$\textrm{H}_2$ }
\newcommand{\HTWOO}{$\textrm{H}_2 \textrm{O}$ }
\newcommand{\CO}{$\textrm{CO}$ }

\newcommand{\kms}{\textrm{ km } \textrm{s}^{-1}}
\begin{document}

   \title{A search for TiO in the optical high-resolution transmission spectrum of HD 209458b: Hindrance due to inaccuracies in the line database}
   	\titlerunning{A search for TiO in the transmission spectrum of HD 209458b}


   \author{H.J. Hoeijmakers
          \inst{1}
          \and
          R.J. de Kok \inst{1,}\inst{2}
          \and
          I.A.G. Snellen
          \inst{1}
          \and
          M. Brogi
          \inst{1} \fnmsep\thanks{NASA Hubble Fellow}
            \and
          J.L. Birkby
          \inst{1,}\inst{3} \fnmsep\thanks{NASA Sagan Fellow}
             \and
          H. Schwarz
          \inst{1}
          }

   \institute{Leiden Observatory, University of Leiden,
              Niels Bohrweg 2, 2333CA Leiden, The Netherlands\\
              \email{hoeijmakers@strw.leidenuniv.nl}
         \and
         	SRON, Netherlands Institute for Space Research, Sorbonnelaan 2, 3584 CA Utrecht, The Netherlands
         \and
             Harvard-Smithsonian Center for Astrophysics, 60 Garden Street, Cambridge MA 02138, USA
             }

   \date{Received August 12, 2014; accepted October 28, 2014}


  \abstract
  {The spectral signature of an exoplanet can be separated from the spectrum of its host star using high-resolution spectroscopy. During such observations, the radial component of the planet's orbital velocity changes, resulting in a significant Doppler shift which allows its spectral features to be extracted.}
   {In this work, we aim to detect TiO in the optical transmission spectrum of HD 209458b. Gaseous TiO has been suggested as the cause of the thermal inversion layer invoked to explain the dayside spectrum of this planet.}
  {We used archival data from the 8.2m Subaru Telescope taken with the High Dispersion Spectrograph of a transit of HD209458b in 2002. We created model transmission spectra which include absorption by TiO, and cross-correlated them with the residual spectral data after removal of the dominating stellar absorption features. We subsequently co-added the correlation signal in time, taking into account the change in Doppler shift due to the orbit of the planet.}
  {We detect no significant cross-correlation signal due to TiO, though artificial injection of our template spectra into the data indicates a sensitivity down to a volume mixing ratio of $\sim 10^{-10}$. However, cross-correlating the template spectra with a HARPS spectrum of Barnard's star yields only a weak wavelength-dependent correlation, even though Barnard's star is an M4V dwarf which exhibits clear TiO absorption. We infer that the TiO line list poorly match the real positions of TiO lines at spectral resolutions of $\sim 100,000$. Similar line lists are also used in the PHOENIX and Kurucz stellar atmosphere suites and we show that their synthetic M-dwarf spectra also correlate poorly with the HARPS spectra of Barnard's star and five other M-dwarfs. We conclude that the lack of an accurate TiO line list is currently critically hampering this high-resolution retrieval technique.}
   {}


   \maketitle
%

\section{Introduction}


\subsection{High Dispersion Spectroscopy}
\begin{figure}
   \centering
   \includegraphics[width=9cm]{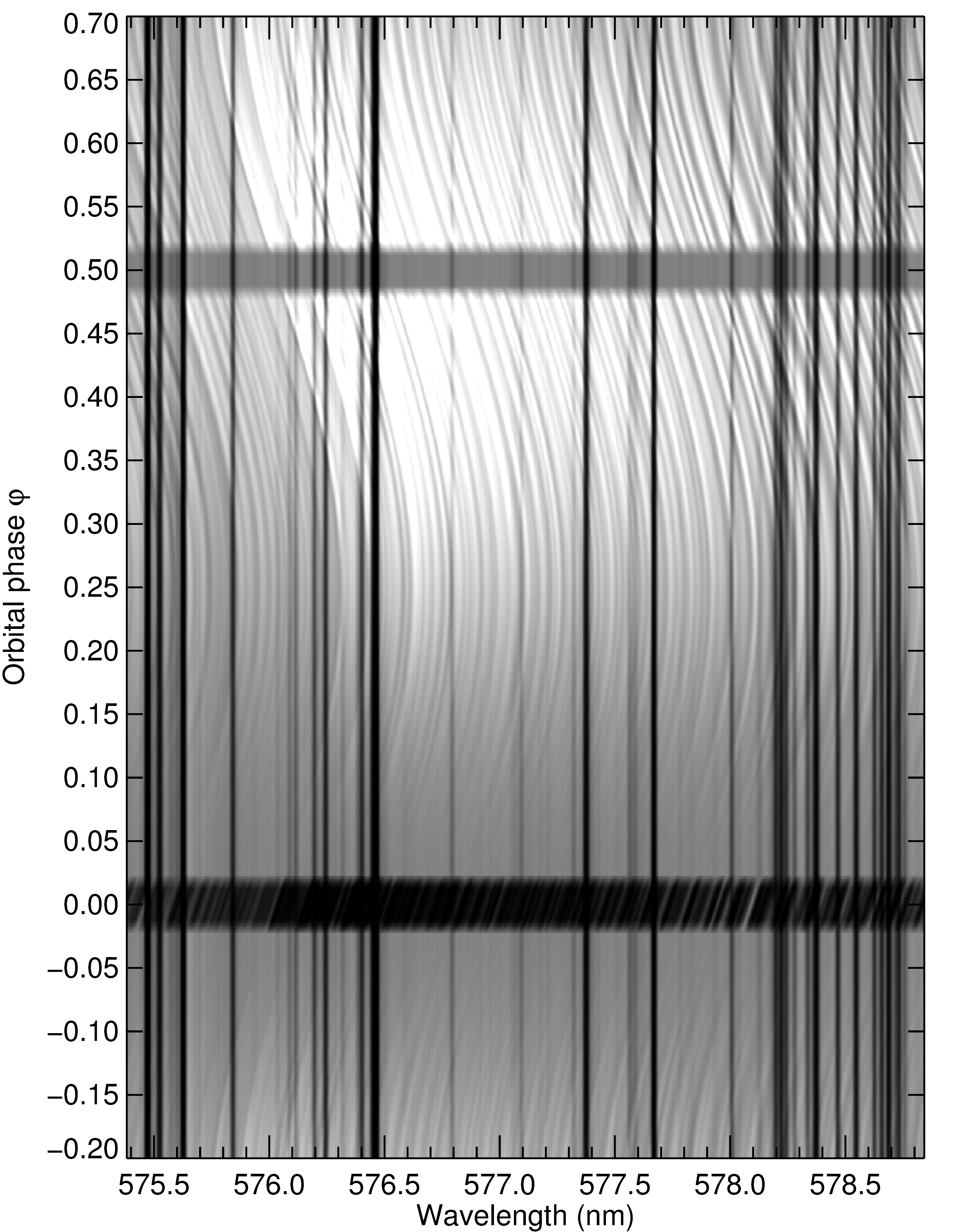}
      \caption{Toy model of the phase-dependent Doppler-shift of TiO lines along the orbit of HD209458b. The white curves represent TiO-emission features due to the inversion layer (greatly enhanced for visual purposes). During transit around $\phi = 0$, TiO produces slanted absorption lines (black). The black vertical lines are stellar absorption lines which are stable in time. This difference in the behaviour of stellar and planetary features provides a means of contrast between star and planet.}
         \label{toymodel}
\end{figure}

Recently, carbon monoxide was identified in the transmission spectrum of the exoplanet HD209458b \citep{Snellen2010}. During transit, the radial velocity of the hot-Jupiter changes by a few tens of $\kms$, resulting in a changing Doppler shift which is measurable at a resolution of $R \sim 10^5$. This time-variation of the planet's spectral signature provides a powerful tool to discriminate between the planet and the overwhelmingly bright host star, because the absorption features in the stellar spectrum are quasi-invariant on these time scales (see Figure \ref{toymodel}). Futhermore, because the orbital period of the planet is well known, its radial velocity is known at all times and the Doppler shift of the planets' absorption features can be targeted specifically.

This technique has also been successfully applied during other parts of of the orbit, where direct thermal emission of the planet is probed. Absorption in the dayside spectra of both transiting and non-transiting hot-Jupiters has been observed: \CO and \HTWOO absorption in the atmospheres of $\tau$ Bo\"otes b, HD 189733 b, 51 Peg b and HD 179733 b \citep{Birkby2013,Brogi2012,Brogi2013,Brogi2014,DeKok2013,Lockwood2014,Rodler2013}. In case of the non-transiting planets, this has allowed for their previously unknown orbital inclinations to be determined: $44.5^{\circ} \pm 1.5^{\circ}$ for $\tau$ Bo\"otes b \citep{Brogi2012}, consistent with the measurement of \citet{Rodler2012}, $79.6^{\circ} - 82.2^{\circ}$  for 51 Peg b \citep{Brogi2013}, and $67.7^{\circ} \pm 4.3^{\circ}$ for HD 179733 b \citep{Brogi2014}, which in turn lead to a determination of their masses.

	\subsection{Inversion layer in HD 209458b}
	HD 209458b is a well-studied exoplanet. It orbits a solar-type star once every $3.5 \days$, located $47 \pc$ from the Sun (see Table \ref{tab:systemparameters} for the physical properties of the system). Using the Spitzer Space Telescope \citet{Knutson2008} found evidence for $\textrm{H}_2 \textrm{O}$ emission in the IRAC bands at $4.5 \um$ and $5.8 \um$, indicative of a thermal inversion layer in the atmosphere of the planet. A temperature inversion is caused by the absorption of incident starlight in a high-altitude layer of the atmosphere. On Earth, solar UV radiation is absorbed in the oxygen-ozone cycle, effectively absorbing all sunlight short of $350 \nm$, heating the atmosphere and forming the stratosphere between $10$ and $50 \km$ altitude \citep[see e.g.][and references therein]{Portmann2007}. In the atmospheres of hot-Jupiters, similar processes may occur, albeit in different chemical environments.
	
A number of compounds have been proposed to be capable of causing the inversion layer in the atmosphere of HD 209458b, one of which is TiO \citep{Hubeny2003, Burrows2007}, a diatomic molecule which is known to cause major absorption features throughout the optical and near-infrared spectra of M-dwarfs. An atmosphere containing a strong optical absorber like TiO would furthermore be consistent with HD 209458b's low albedo \citep{Rowe2006}. \citet{Desert2008} reported a tentative detection of TiO absorption in the transmission spectrum of HD 209458b, though more recent broadband studies seem to support a lack of TiO absorption in the transmission spectra of other hot-Jupiters \citep[see e.g.][]{Huitson2013,Sing2013, Gibson2013, Bento2014}. Also, models of hot-Jupiter atmospheres suggest that maintaining a significant gaseous TiO concentration at high altitudes is difficult due to condensation of TiO either at depth or at the night-side \citep{Spiegel2009, Burrows2009, Fortney2010, Parmentier2013}. More recent observations of the dayside emission spectrum of HD 209458b, call into question the existence of the inversion layer altogether \citep{Zellem2014, Diamond-Lowe2014}. Interestingly, \citet{Stevenson2014} observed molecular absorption features in the near-infrared dayside spectrum of Wasp-12b, possibly attributable to TiO. This contrasts with the optical transmission spectra of the same planet, put forth by \citet{Sing2013}. Thus, it is as yet unclear whether TiO is an important chemical component of hot-Jupiter atmospheres. 

In this paper we apply a cross-correlation-based retrieval method to search for TiO in the transmission spectrum of HD 209458b at high spectral resolution. Section 2 describes our dataset and the application of the method. Section 3 discusses our results, followed by our conclusions in Section 4.


\begin{table} 
\begin{tabularx}{0.5\textwidth}{|l|l|l|}
\hline
\textbf{Parameter} & \textbf{Symbol} & \textbf{Value} \\ 
\hline

Visible magnitude & $V$ & $7.67 \pm 0.01$ \\ 
Distance (pc) & $d$ & $47.4 \pm 1.6$ \\
Effective temperature (K) & $T_{\textrm{eff}}$ & $6065 \pm 50$ \\
Luminosity ($L_{\odot}$) & $L_*$ & $1.622^{+0.097}_{-0.10}$ \\
Mass ($M_{\odot}$) & $M_*$ & $1.119 \pm 0.033$ \\
Radius ($R_{\odot}$) & $R_*$ & $1.155^{+0.014}_{-0.016}$ \\
$\textrm{Systemic velocity } (\textrm{km}\textrm{s}^{-1})^{a}$ & $\gamma$ & $-14.7652 \pm 0.0016$ \\
Metallicity (dex) & $\left[ \textrm{F}/ \textrm{H} \right]$  & $0.00 \pm 0.05$ \\
Age (Gyr)& & $3.1^{+0.8}_{-0.7}$ \\ 

\hline

\textbf{$\textrm{Orbital period (days)}^{b}$} & $P$             & $3.52474859$  \\
											&					& $\pm 0.00000038$ \\
Semi-major axis (AU)              			& $a$               & $0.04707^{+0.00046}_{-0.00047}$ \\
Inclination (deg)                  			& $i$               & $86.71 \pm 0.05$ \\
$\textrm{Eccentricity}^{c}$					& $e \cos \omega$   & $0.00004 \pm 0.00033$ \\
Impact parameter                   			& $b$               & $ 0.507 \pm 0.005 $ \\
$\textrm{Transit central time (HJD)}^{b}$  	& $t_c$			    & $2,452,826.628521$ \\ 
											&					& $\pm 0.000087$ \\
$\textrm{Transit duration (min)}^{d}$      	& $t_T$             & $183.89 \pm 3.17$ \\
Mass ($M_J$)                               	& $M_p$             & $0.685^{+0.015}_{-0.014}$ \\
Radius ($R_J$)                             	& $R_p$             & $1.359^{+0.016}_{-0.019}$ \\
Density ($\textrm{g }\textrm{cm}^{-3}$)    	& $\rho_p$          & $0.338^{+0.016}_{-0.014}$ \\
Equilibrium temperature (K)                	& $T_{\textrm{eq}}$ & $1449 \pm 12$ \\
\hline
\end{tabularx}
\caption[lalaaa]{Properties of HD 209458 (upper part) and HD 209458b (lower part), adopted from \citet{Torres2008}.\\ 
$a:$ Adopted from \citet{Mazeh2000}.\\ 
$b:$ Adopted from \citet{Knutson2007}.\\
$c:$ Adopted from \citet{Crossfield2012}.\\
$d:$ Adopted from the Exoplanet Orbit Database \citep{Wright2011}.} 
\label{tab:systemparameters}
\end{table}

\section{Observations and data reduction}
\subsection{Subaru data of HD 209458}\label{sec:HDS}
The data used in this analysis were taken on October 25, 2002 using the High Dispersion Spectrograph (HDS) on the Subaru 8.2m telescope. The observations were performed by \citet{Narita2005} and were originally aimed at detecting absorption of sodium and a number of other atomic species in the transmission spectrum of HD 209458b. These were later used by \citet{Snellen2008} who did indeed detect sodium. The data consist of 31 exposures with a mean exposure time of $500 \sec$ which cover one 3h transit, plus 1.5h and 0.5h of baseline before and after, respectively. The spectrum is observed over 20 echelle orders, covering a wavelength range between $554 \nm$ and $682 \nm$ at a spectral resolution of $R \sim 45,000$ ($6.7 \kms$ resolution). Each spectral order is sampled at $0.9 \kms$ per pixel over 4100 pixels. The basic data reduction (such as bias subtraction, flat fielding and 1D spectral extraction) were performed by \citet{Snellen2008} and are described therein in detail.

\subsection{HARPS data of M-dwarfs}\label{sec:HARPS}
\begin{figure*}
   \centering
   \includegraphics[angle=270,width=\linewidth]{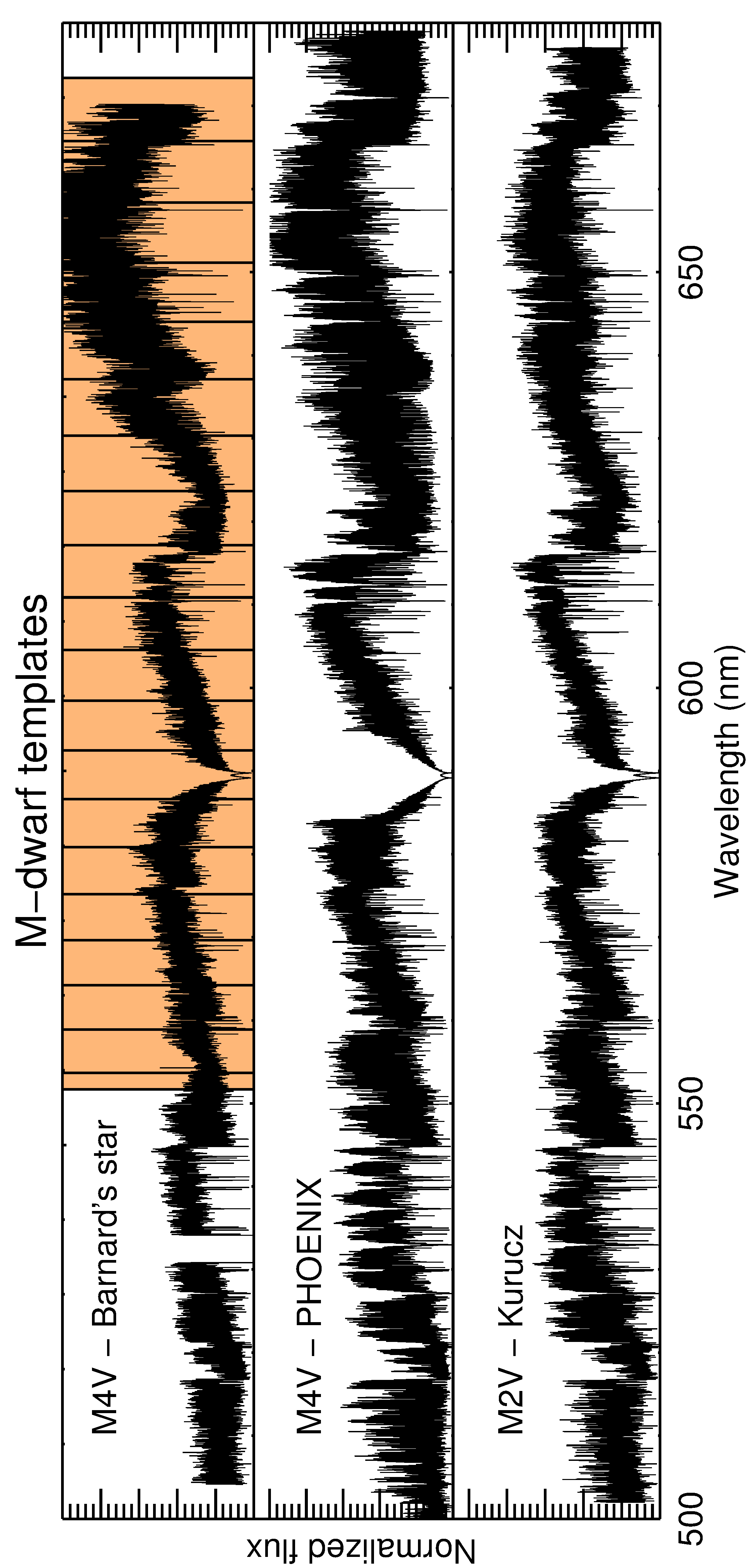}
   \caption{Top panel: Archival HARPS spectrum of Barnard's star at a resolution of $R=115,000$. Middle panel: A PHOENIX model of a generic M4V dwarf. Bottom panel: Kurucz model of Gliese 411, a standard M2V dwarf. These three templates contain strong broadband features, which are removed by applying a high-pass filter later in the analysis (not shown). The orange region in the top panel covers the approximate extent of the spectral orders of the HDS Subaru dataset, which are delimited by the solid vertical lines.}%
\label{templates}
\end{figure*}

Our analysis relies on cross-correlating the transmission spectra of HD 209458b with high-resolution template spectra of TiO-bearing models of HD 209458b's absorption spectrum which rely on line lists published by \citet[see sections \ref{Postproc} and \ref{sec:models}]{Freedman2008}. The optical spectra of M-dwarfs are dominated by TiO absorption features and are therefore natural benchmarks against which we tested the accuracy of these TiO templates. Many nearby M-dwarfs have been observed extensively at high spectral resolution in order to detect the radial velocity signature induced by potential orbiting companions. We downloaded high resolution spectra of 5 different M-dwarfs from the ESO data archive. These spectra were obtained by the HARPS spectrograph at ESO's 3.6m La Silla telescope, between $500\nm$ and $691\nm$ at resolutions of $R=115,000$. Information about the 5 stars is shown in Table \ref{tab:Mdwarfs}. Barnard's star is used throughout this work as the primary example, and is shown in the top panel of Figure \ref{templates} before the removal of all broadband features using a high-pass filter (the spectra of the other M-dwarfs were filtered in the same way), and after correcting for its radial velocity of $-110.51 \kms$ \citep{Nidever2002}.

\begin{table} 
\begin{tabularx}{0.5\textwidth}{l|l|l|l}
\textbf{Name} & \textbf{Spectral type} & \textbf{PID (ESO)} &\textbf{PI} \\ 
\hline
Proxima Centauri & M6.0V & 183.C-0437(A) &	Bonfils \\
GJ402 & M5.0V & 183.C-0437(A) & Bonfils \\
GJ9066 & M4.5V & 072.C-0488(E) & Mayor	\\
GJ876 & M5.0V & 183.C-0437(A) &	Bonfils \\
Barnard's star & M4.0V & 072.C-0488(E) & Mayor \\

\end{tabularx}
\caption[lalaaa]{Overview of the archival M-dwarf spectra used in this work, obtained using the HARPS-south instrument.}  
\label{tab:Mdwarfs}
\end{table}

\subsection{Post-processing and cross-correlation}\label{Postproc}
The analysis of the high dispersion transmission spectra of HD 209458b differs at several points from that of infrared data used in previous high dispersion spectroscopy studies by our group \citep[see][]{Birkby2013, Brogi2012, Brogi2013, Brogi2014, DeKok2013}, which are dominated by time-dependent telluric absorption features, mainly due to variations in airmass. In the optical, the atmosphere presents no significant absorption features except in some specific wavelength ranges. Rather, the spectrum is dominated by stellar absorption lines. Assuming that the stellar spectrum is stable during the transit, it can be removed by dividing each of the individual spectra in the time-series with the time-averaged stellar spectrum. In principle, this operation does not affect the planet signal because it significantly shifts in wavelength during the observations due to the change in the radial component of the orbital velocity during transit (from $-15 \kms$ to $+15 \kms$).

\begin{figure*}
   \centering
   \includegraphics[angle=270,width=\linewidth]{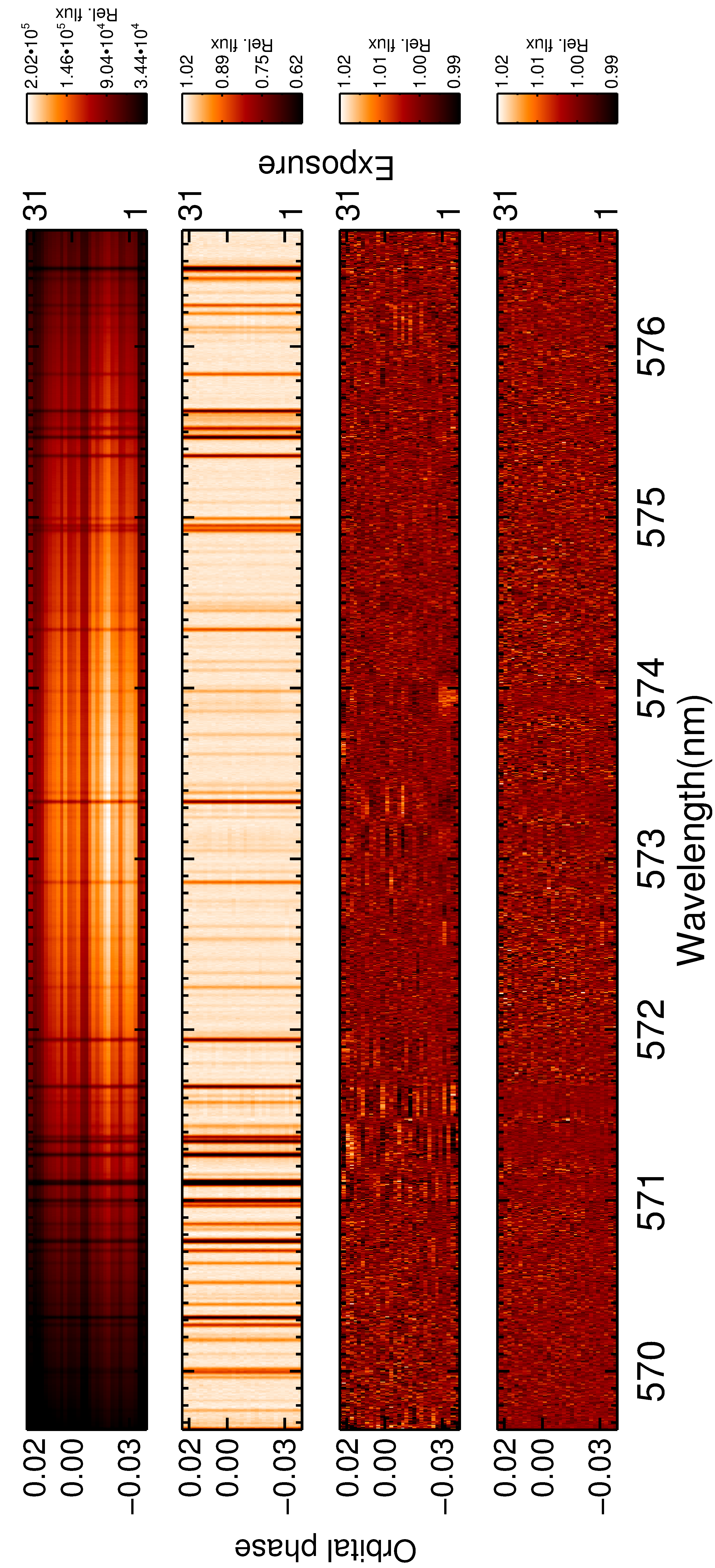}
   \caption{The major post processing steps of echelle order 17 before cross-correlation. The top panel shows the initial state of the spectral order, arranged such that all exposures in this order are stacked vertically. The result of the removal of the blaze pattern and the alignment of the individual exposures (which cannot be seen by eye as the misalignments are on a sub-pixel level) is shown in the second panel. The third panel shows the residuals after dividing through the time-averaged spectrum (i.e. the average of the spectrum obtained by averaging each column in the second panel). The fourth panel shows the same residuals after removing $4 \sigma$ outliers, removing latent broadband features and weighing down noisy columns. The average standard deviation of the residuals at this stage is $0.45\%$. Note that only the lower two panels are scaled to the same color range.}%
\label{arrays}
\end{figure*}

The full sequence of processing steps is summarized below. Intermediate data products are shown in Figure \ref{arrays}.
	
	\begin{enumerate}
	\item[1] \textbf{Broadband removal and normalization:} To remove strong broadband components such as the stellar continuum and the spectrograph's blaze function, the brightest exposure of each spectral order was divided into 50 wavelength regions. A 7-th order polynomial was fit through the maximum of each part to obtain the continuum level. This fit was removed from all other exposures through division, removing the strongest broadband features. 	
	Subsequently, the continuum of each of the exposures was obtained by dividing each exposure into the same 50 regions as above and again finding the maximum of each part. These continua were interpolated, and removed through division. This acts to remove residual broad-band variations not removed by the 7th-order polynomial fit, and normalizes the continuum of each exposure to unity. At this stage there were still significant residual broad-band variations, which were removed at a later stage (see step 7).
	\item[2] \textbf{Alignment of spectra:} The spectra slowly drift in wavelength due to the change in radial velocity of the observatory during the observations, and instrumental instability. These misalignments are of the order of a pixel and would adversely affect the removal of the time-average stellar spectrum if left untreated. In each spectral order a dozen strong stellar absorption lines were identified by eye. To these lines in each of the 31 exposures, Gaussian profiles were fitted to obtain the position of each line center. A linear fit through the misalignments of the different absorption lines was used to align the spectra to a common reference frame (a higher order fit was not warranted because the scatter in the determination of the line centroids was too high to identify higher order components).
	\item[3] \textbf{Wavelength solution:} As HD 209458 is a close solar analogue, we matched our time-averaged spectrum (see next step) of the HD 209458 system to a model solar spectrum (adopted from the Kurucz stellar atmosphere atlas) in a step-wise fashion. The strongest stellar absorption lines were identified and matched by eye. This produced a crude wavelength solution which was subsequently used to identify 30 - 40 less prominent lines in each order. We subsequently fitted Gaussian profiles to the cores of these lines in both the data and the solar template, producing a wavelength solution with a standard deviation of $1.7 \times 10^{-3} \nm \pm 0.7 \times 10^{-3}$, corresponding to $0.83 \pm 0.36 \kms$ or $ \sim 12 \%$ of the spectral resolution. Because the solar spectrum is synthesized at rest, the wavelength solution is calibrated to the rest frame of HD 209458 and the mean radial velocity with respect to the observatory on Earth is thus automatically corrected for. Any time-dependent Doppler shifts had already been corrected in step 2. Therefore, each exposure is now in the rest frame of the host star.
		\item[4] \textbf{Removal of stellar spectrum:} Each pixel in each spectral order was time-averaged by taking the median of the aligned spectra, which corresponds to taking the median of each of the columns of the array in panel 2 of Figure \ref{arrays}. This average contains all time-constant stellar features, but only a small part of the time-varying planet signal, since it is shifted to a different column in each exposure. It is removed from each exposure through division. The width of planetary absorption lines is limited to $6.7\kms$, which is the resolution of the instrument. Between exposures, the planet signal is shifted by $\sim1.8 \kms$. Therefore, the planet signal is present in no more than 4 exposures per pixel column. These 4 exposures are part of a median with 27 other exposures which don't contain the planet signal. Therefore, we expect that up to $10\% - 15\%$ of the planet signal is removed along with the median stellar spectrum. This has no importance for the rest of the analysis, because we measure the significance of any TiO detection by our ability to retrieve injected model spectra, which suffer in the same way from this degradation (see section \ref{sec:Results}).
	\item[5] \textbf{Removal of bad pixels and cosmics:} The resulting residuals contain bad pixels and cosmic rays. All pixels deviating by more than $4 \sigma$ from the mean were reset to the mean, removing all significant bad pixels and cosmic rays. 0.21\% of all pixel values were affected.
	\item[6] \textbf{Normalization by signal to noise:} The noise in the residual spectrum varies with pixel position due to the presence of deep stellar absorption lines up until step 4. We divide each pixel in the residual spectrum by the variance of that pixel value (variance of each column of panel 3, Figure \ref{Postproc}) to weigh down such pixels. Low SNR regions would otherwise add a significant amount of noise to the cross-correlation function.
	\item[7] \textbf{Suppression of broadband residuals:}  The residuals feature broadband variations due to imperfect removal of instrumental effects and broad stellar lines. Latent broadband variations were characterized by smoothing each exposure with a Gaussian with a width of 40px ($36 \kms,$) and were divided out of the data. The last panel of Figure \ref{arrays} shows the residual of this final processing step, on which the cross-correlation is performed.
	\item[8] \textbf{Cross correlation:} The planet's potential absorption features were extracted by cross-correlating the residual spectra with template spectra of TiO (see section \ref{sec:models}) across the full wavelength range of the data i.e. between $\sim 554$ and $\sim 682 \nm$. The templates were Doppler-shifted to velocities ranging between $\pm 150 \kms$ relative to the star, in steps of $1\kms$ (which roughly corresponds to the sampling rate of the data, $0.9\kms$ per pixel). At the Doppler shift equal to the radial velocity of the planet, the cross-correlation effectively co-adds all of the individual absorption lines of the molecule.
	\end{enumerate}	
	
\subsection{Model spectra}\label{sec:models}
\begin{figure*}
   \centering
   \includegraphics[width=\linewidth]{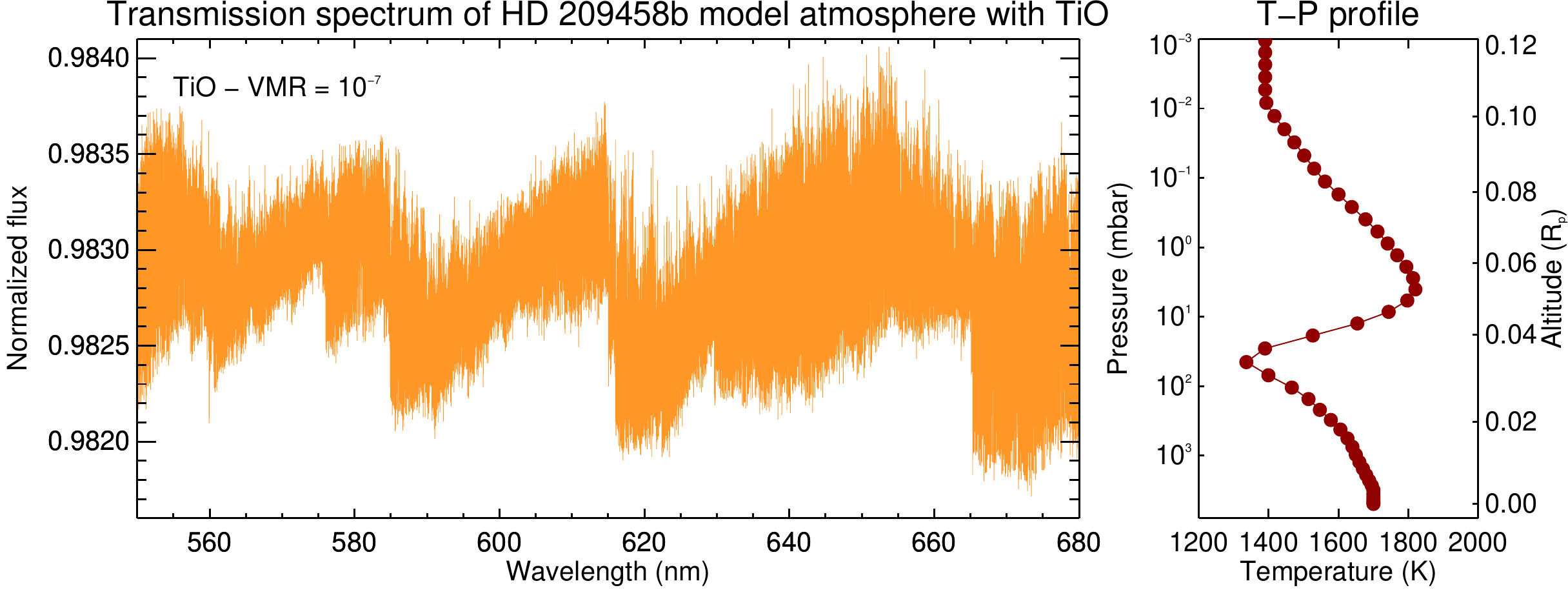}
   \caption{The left panel shows one of our model absorption spectra of HD 209458b containing TiO at a VMR of $10^{-7}$, normalized to the brightness of the system out of transit and according to the temperature-pressure profile shown on the right, along with the corresponding altitude in units of the planet radius. Throughout this work, we employ this spectrum as a template in various cross-correlation analyses, as well as four similar templates at lower VMRs; and one template at a \textbf{constant} temperature of $3000 \Kelvin$. Note that the form of the assumed T-P profile does not significantly impact the correlation, because the effective temperature of the star is $\sim 6065 \Kelvin$.} 
\label{models}
\end{figure*}

For the purpose of applying the cross-correlation method to the case of TiO in the atmosphere of HD 209458b  (see below), we generated several model transmission spectra of atmospheres containing various concentrations of TiO, at a sampling resolution of $R=240,000$. The line database stems from \citet{Freedman2008}, who presented a modified line list based on the one published by \citet{Schwenke1998}. The model atmosphere has a temperature-pressure profile as shown in Figure \ref{models}, and includes \HTWO scattering and molecular absorption due to TiO with volume mixing ratios (VMR) of $10^{-7}$, $10^{-8}$, $10^{-9}$, $10^{-10}$ and $10^{-11}$. The solar photospheric abundance of titanium is $10^{-7.1}$ \citep{Asplund2006}, so the transmission spectrum of TiO is modelled between $\sim 1$ and $\sim 10^{-4}$ times the solar abundance of Ti. These mixing ratios were assumed to be uniform throughout the atmosphere. As such, no chemistry, phase-changes or dynamics were taken into account. The atmosphere was assumed to have a temperature inversion as displayed in the right panel of Figure \ref{models} \citep[adopted from][]{Burrows2007}, but because the background continuum originates from a star with an effective temperature of $6065 \Kelvin$, the assumption of a particular T-P-profile does not significantly influence the template spectrum. To confirm the robustness of the template to a temperature mismatch, we also modelled a TiO-bearing atmosphere at a constant temperature of $3000 \Kelvin$ and cross-correlated it with the template shown in Figure \ref{models}. The result is shown in Figure \ref{xcorcompare}, and discussed in more detail in section \ref{sec:evaluation}.

For benchmark purposes (see section \ref{sec:HARPS}), we also cross-correlated with synthetic M-dwarf spectra originating from the Kurucz and PHOENIX stellar atmosphere model suites which necessarily include models of the TiO molecule. Both models use a line list calculated by \citet{Plez1998}. The Kurucz model is a template spectrum of the M2V dwarf GL411 at a resolution of $R=100,000$ \citep{Castelli2003} and the PHOENIX model, a template of a generic M4V dwarf ($T_{\textrm{eff}}=3100\textrm{K}$, $\log g = 5.0$, $\textrm{Fe/H} = +0.5$, obtained from the PHOENIX online database as presented by \citet{Husser2013}) at a resolution of $R=125,000$. They are shown together with the HARPS spectrum of Barnard's star in Figure \ref{templates}.

\section{Results and discussion}\label{sec:Results}
\begin{figure*}
   \centering
   \includegraphics[angle=270,width=\linewidth]{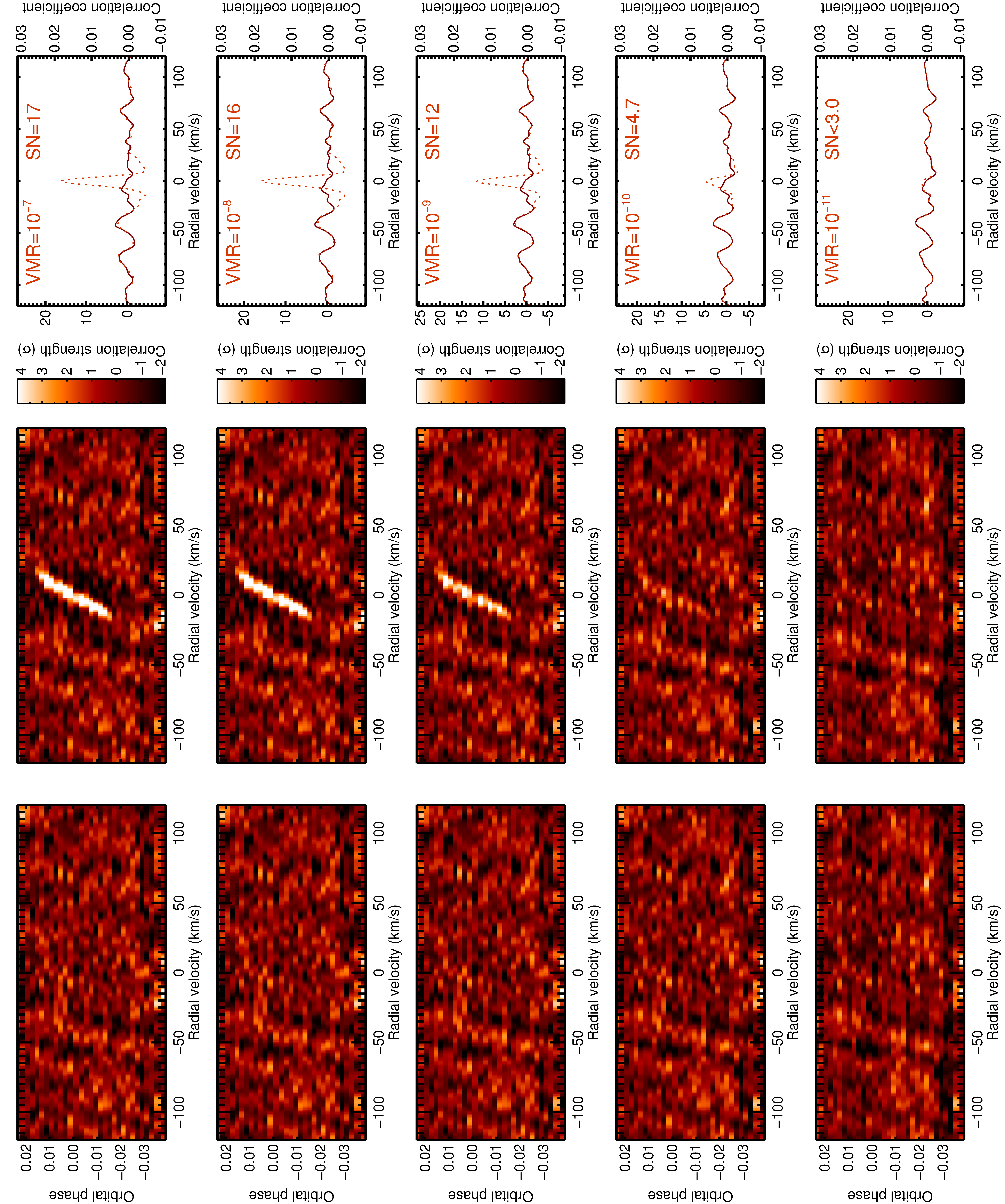}
   \caption{The strength of the cross-correlation signature as a function of radial velocity and orbital phase. Each row of panels shows the cross-correlation with model atmospheres with different VMR's of TiO (left panels, ranging from $10^{-7}$ (top) to $10^{-11}$ (bottom) by factors of 10). The models are also injected into the data prior to cross-correlation (middle panels) to probe the sensitivity of the procedure. Co-addition of all exposures (right panels) is achieved by shifting each exposure (i.e. each row in the middle and left panels) to the rest-frame of the planet and averaging them in time. The dashed lines are the co-added correlation strength of the residuals, after having injected them with the model template with which the cross-correlation is performed. The solid lines are the co-added correlation strengths of the residuals without prior injection of the template. The signal-to-noise of the correlation peak is calculated by taking the peak correlation, and dividing it through the standard deviation of the cross-correlation response, excluding a range of $\pm 20 \kms$ in which the correlation peak itself resides. Clearly, the presence of TiO would be convincingly retrieved at the ppb level, provided that the line database is accurate.}
   
\label{xcorresult}
\end{figure*}

The left panels of Figure \ref{xcorresult} show the correlations between the HDS spectra of HD 209458b and our TiO template at five modeled VMRs (see section \ref{sec:models}). The correlation coefficients are plotted as a function of the radial velocity to which the template is shifted (horizontal axis) for every individual exposure (i.e. time, vertical axis).

To assess the ability of the cross-correlation procedure to retrieve the template spectra from real data, we inject the templates into the data prior to cross-correlation using the known orbital velocity of HD 209458b, between steps 3 and 4 in section \ref{Postproc}, similar to \citet{Snellen2010}. Prior to injection, we convolve the template with a Gaussian to match the spectral resolution of the data $(45,000)$, and convolve again with a box-function to account for the widening of the planetary absorption lines due to the changing Doppler-shift during the $500 \textrm{s}$ exposures. This gives rise to the middle  panels in Figure \ref{xcorresult}. Injection of a TiO-bearing template at a VMR greater than $10^{-10}$ clearly produces a slanted feature, the gradient of which is a measure of the orbital velocity of the planet at which the template was injected. 

Given that the radial velocity of the planet is known at all times, each exposure (i.e. row in Figure \ref{xcorresult}) can be shifted to the rest-frame of the planet. This allows for co-addition of all exposures, the result of which is shown in the right panels in Figure \ref{xcorresult}. Here, the solid lines represents the correlation of the templates with the residuals of the data, and the dashed line represents the correlation after injection of the template. The signal-to-noise of the observed correlation peak is indicative of the significance at which the template would be retrieved, had it been  identically present in the residuals. Evidently, the peak correlation strength decreases with decreasing VMR, hence the VMR at which the correlation peaks at the $3 \sigma$ level, we consider to be the theoretical limiting VMR at which TiO would be retrievable by applying the current analysis on this data. 

It is clear from the left panels of Figure \ref{xcorresult} that the residuals do not correlate with our TiO templates. This non-detection is significant down to VMRs of $10^{-10}$, as evidenced by the successful retrieval of our templates after having artificially injected them into the data. This sensitivity limit corresponds to $10^{-3}$ times the solar abundance of Ti \citep{Asplund2006}. The models of the atmosphere of HD209458b are greatly simplified approximates to the real physical environment. Especially the assumption of a uniform VMR across the entire atmosphere is questionable, as the atmosphere is not expected to be homogeneous. However, by extending the absorbing TiO layer uniformly throughout the atmosphere, the TiO absorption at that VMR is maximized. The limiting VMR of $10^{-10}$ is therefore a true lower limit: If TiO would only be present at a VMR of $10^{-10}$ in parts of the atmosphere, it would not be retrieved in this analysis.

Before invoking a plethora of reasons for this non-detection of TiO such as hazes, obscuring clouds, screening by other species, cold-traps or destructive chemistry, it should be noted that the sensitivity of the cross-correlation procedure critically depends on the accuracy of the line positions in the template spectrum. If the wavelength positions of the lines are inaccurate, the cross-correlation procedure may not add their contributions constructively. As shown below, this is a real concern. 

\subsection{Evaluation of template spectra}\label{sec:evaluation}
Because TiO absorption visibly dominates M-dwarf optical spectra, a high-resolution optical spectrum of Barnard's star can be used to test the accuracy of our TiO template. Their mutual correlation should be very clear.

\begin{figure}
   \sidecaption
   \includegraphics[angle=270,width=9cm]{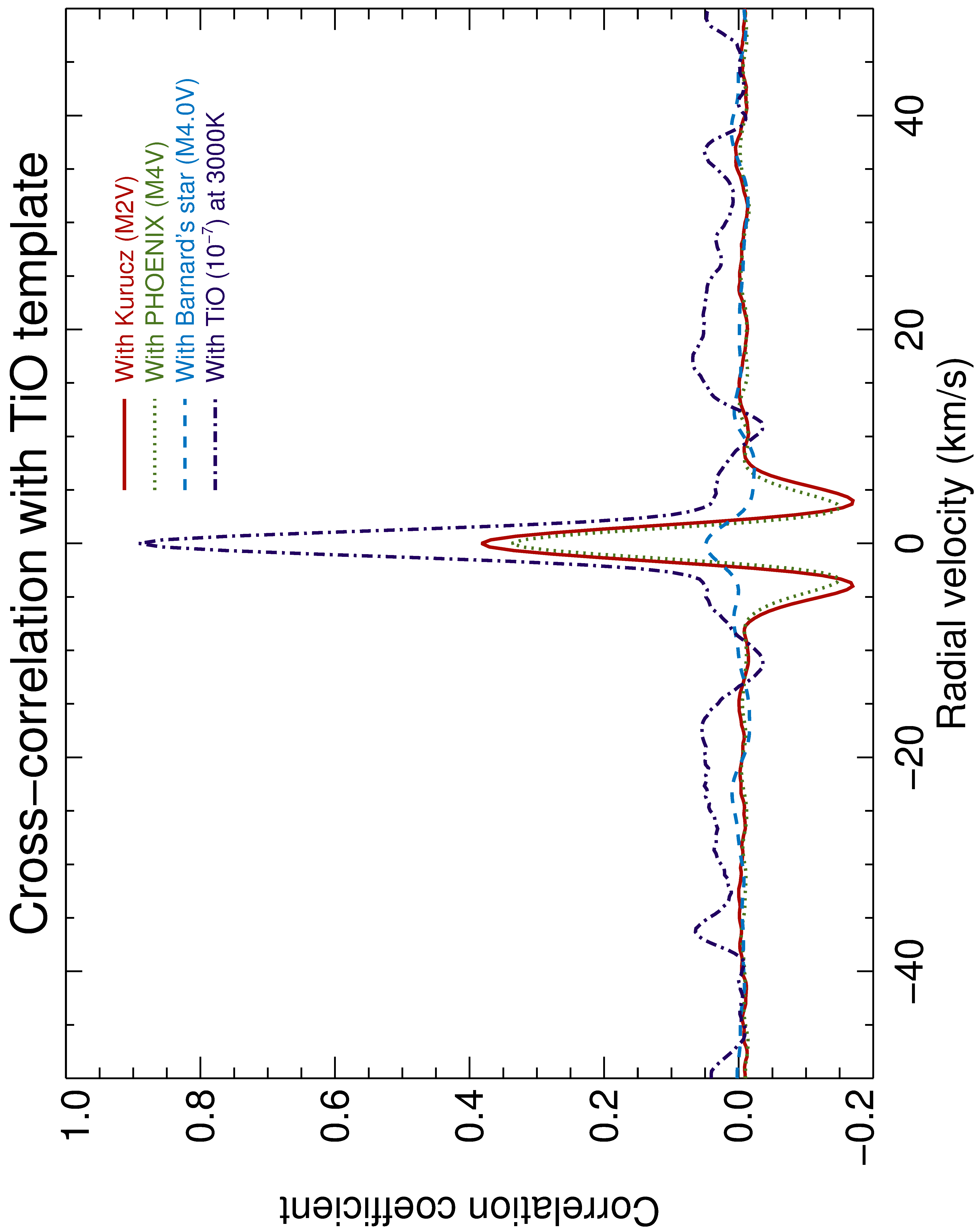}
   \caption{Correlation of our TiO template (at $T\sim1500\Kelvin$), as shown in Figure \ref{models}, with the PHOENIX M4V model (green), the Kurucz model of GL411 (red), Barnard's star's optical spectrum (light blue) and the TiO template modelled at a temperature of $3000\Kelvin$ (dark blue). All modelled spectra correlate strongly with our TiO template, while Barnard's star's spectrum does not. The high peak correlation between the two templates at different temperatures suggests that the assumption of an inaccurate T-P profile has little effect on the templates.}
\label{xcorcompare}
\end{figure}

Figure \ref{xcorcompare} shows the cross-correlation between our TiO template at different temperatures, the stellar atmosphere models and the HARPS-spectrum of Barnard's star. The TiO-bearing model of HD 209458b (VMR=$10^{-7}$, T-P profile as shown in Figure \ref{models}) shows a correlation peak of 0.35 with the stellar atmosphere models. The cross-correlation between our TiO templates at $3000\Kelvin$ and $\sim 1500\Kelvin$ peaks at 0.9. M-dwarf transmission spectra feature a host of other optical absorbers, which decrease the correlation between an M-dwarf spectrum and a pure-TiO template. We regard this as the main reason for the non-prefect correlation between our TiO template and the stellar atmosphere models, rather than a mismatch of the T-P profile associated with the template.

In contrast with the stellar atmosphere models, a marginal correlation peak of only $0.1$ is observed between the TiO template and the HARPS spectrum. From this we may conclude that our TiO-bearing model spectra do not accurately reproduce real-world TiO absorption at high resolution. 

\begin{figure}
   \includegraphics[width=\linewidth]{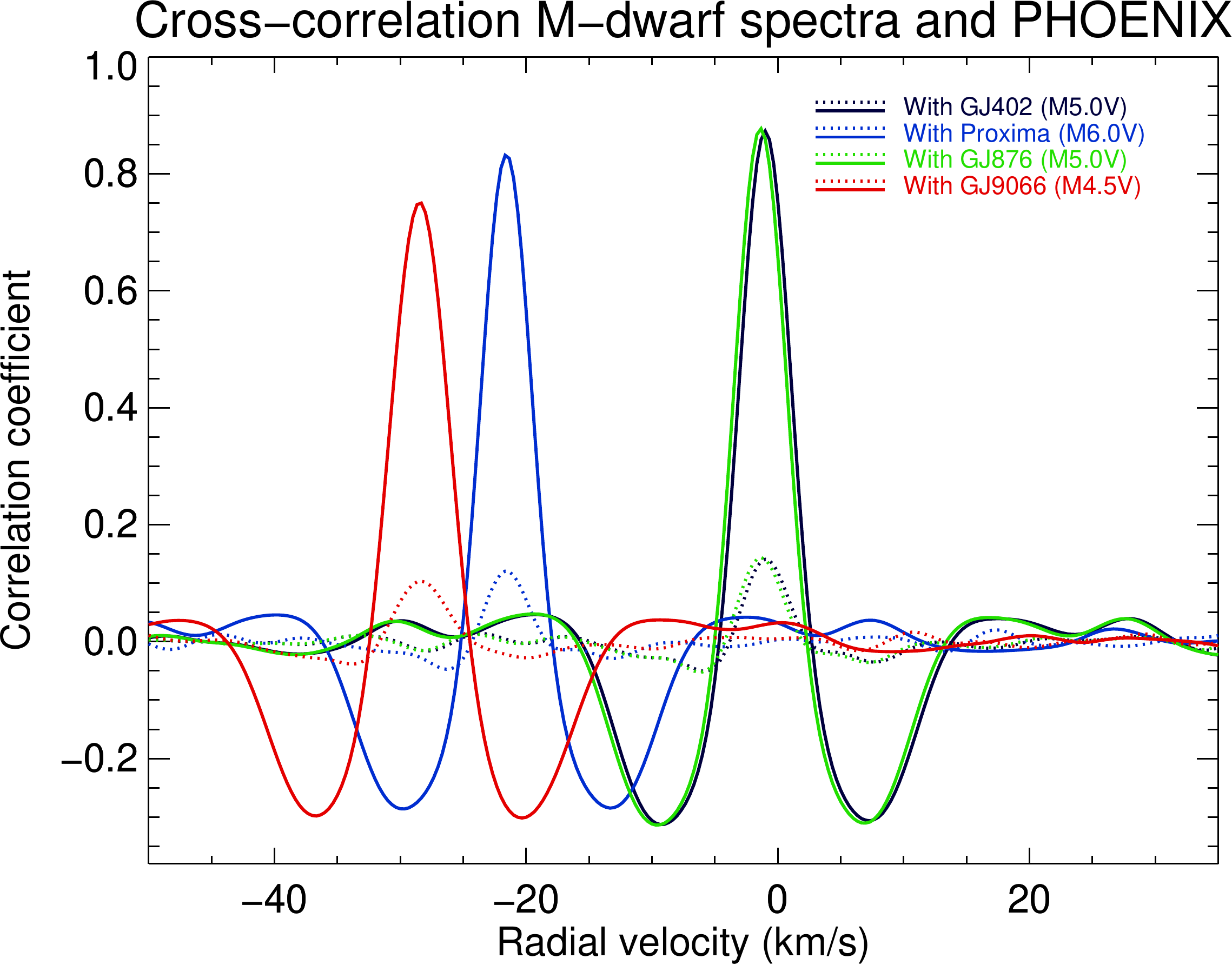}
   \caption{Correlation of several M-dwarf spectra (see Table \ref{tab:Mdwarfs}) with Barnard's star (solid lines), and with the PHOENIX M4V model (dashed lines). The cross-correlation procedure retrieves the Doppler shifts of the other M-dwarfs, as we did not shift these to their own rest frames, for clarity. All M-dwarfs correlate by more than 0.75 with Barnard's star, while correlation with the model absorption spectrum never exceeds 0.15.} %
\label{Mdwarf_mutual}
\end{figure}

\begin{figure*}
   \centering
   \includegraphics[angle=270,width=0.75\linewidth]{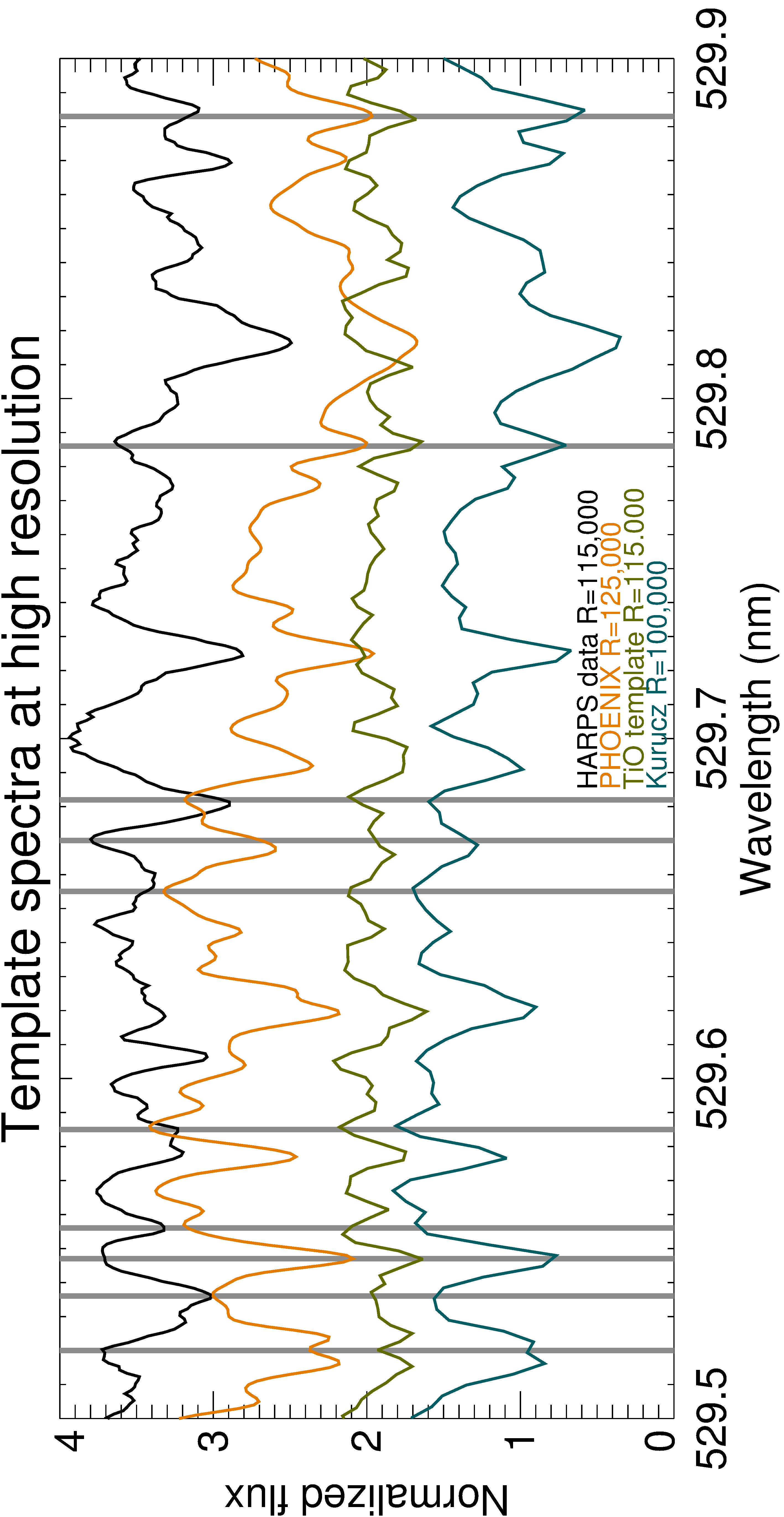}
   \caption{A visual comparison of the three TiO-bearing model atmospheres and the HARPS-spectrum of Barnard's star. The spectra are vertically offset for clarity. Although some broad absorption lines of the stellar models seem to line up with the HARPS-data, most narrow-band features (the majority of them are due to TiO) are clearly misplaced. Some obvious mismatches are indicated by the gray vertical lines.}%
\label{resolution-compare}
\end{figure*}

To test the extent at which stellar atmosphere models and real M-dwarf spectra match each other at high resolution, we cross-correlate Barnard's star's spectrum with the spectra of the other 5 M-dwarfs, and with the PHOENIX M4V model. This is shown in Figure \ref{Mdwarf_mutual}. Regardless of differences in effective temperature, metallicity and signal-to-noise, real M-dwarf spectra correlate strongly with each other; with a peak of >0.75 in all cases. However, as is the case with the TiO template, the PHOENIX model correlates at less than 0.15 with all of the M-dwarf spectra. This strengthens our conclusion that the modelled spectrum of TiO is not representative of real TiO. The high correlation between the TiO template and the PHOENIX and Kurucz models, suggests that the TiO line lists have a common source, and the same uncertainties in the line positions seem to have persisted in the works of both \citet{Plez1998} and \citet{Schwenke1998}. A visual comparison of the models and the spectrum of Barnard's star, shows great similarity between all three models at high resolution, but a severe mismatch with the HARPS data (Figure \ref{resolution-compare}).

\begin{figure*}
   \sidecaption
   \includegraphics[angle=0,width=12cm]{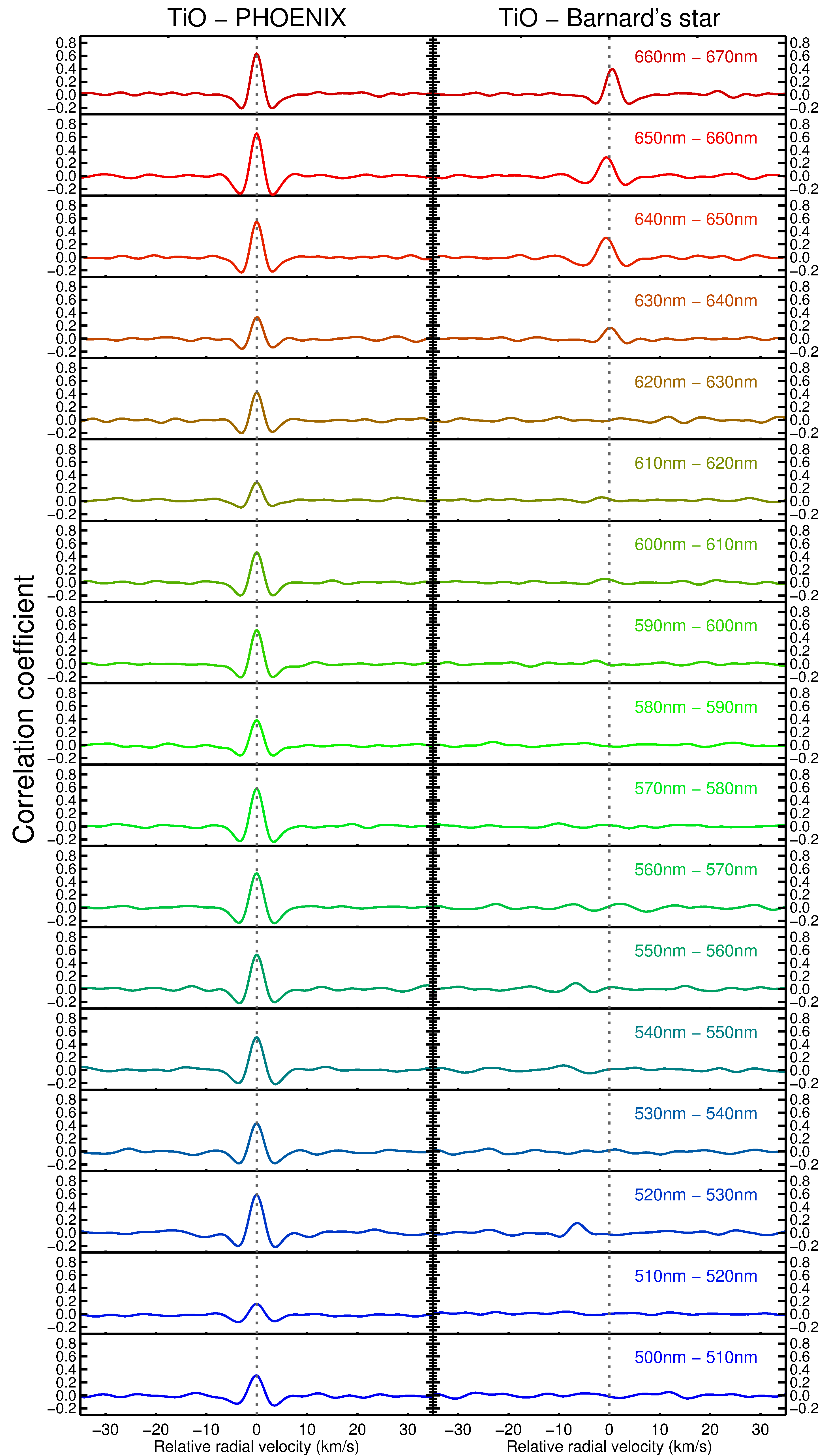}
   \caption{Cross-correlation of sections of the TiO template with the PHOENIX model (left column) and with Barnard's star's spectrum (right column), at different wavelengths. The TiO template correlates strongly almost everywhere between $500 \nm$ and $670 \nm$, while the correlation between the template and the TiO-dominated spectrum of Barnard's star is absent across most of the wavelength range, and less than $0.4$ in the best case. We also note that maximum correlation occurs at slightly different radial velocities (an extreme feature is observed between $520 \nm$ and $530 \nm$). This is explicitly indicative of wavelength-offsets of entire TiO bands (and thus inaccuracies in the determination of the energy levels of the TiO molecule).}%
\label{xcor-ladder}
\end{figure*}

Although the TiO models do not match the HARPS data well, our TiO template is able to partially retrieve the presence of TiO in Barnard's star spectrum. The extent of the inaccuracies of the models as shown in Figure \ref{resolution-compare} near $530\nm$, raises the suspicion that the marginal correlation with Barnard's star is not due to the haphazard alignment of model absorption lines. Instead, it is possible that certain absorption bands are + more accurately than others. To test the hypothesis that inaccuracies are associated with certain absorption bands only, we divide the available $130 \nm$ range of our HARPS data into smaller sections and again cross-correlate our model of HD 209458b with each section separately. The result is shown in Figure \ref{xcor-ladder}.

Over most of the wavelength range, our model does not correlate with the TiO spectrum. However above $630 \nm$, the TiO template appears to better match the data, increasing in correlation towards longer wavelengths (a trend which may continue further towards the near infra-red). The four sections shown in in Figure \ref{xcor-ladder} correlate maximally at slightly different radial velocity shifts (few km/s), meaning that ensembles of lines are collectively offset. This specifically indicates inaccuracies in the calculation of the energy levels of the TiO molecule. 


\subsection{Alternative template spectra}
\begin{figure*}
   \centering
   \includegraphics[angle=270,width=\linewidth]{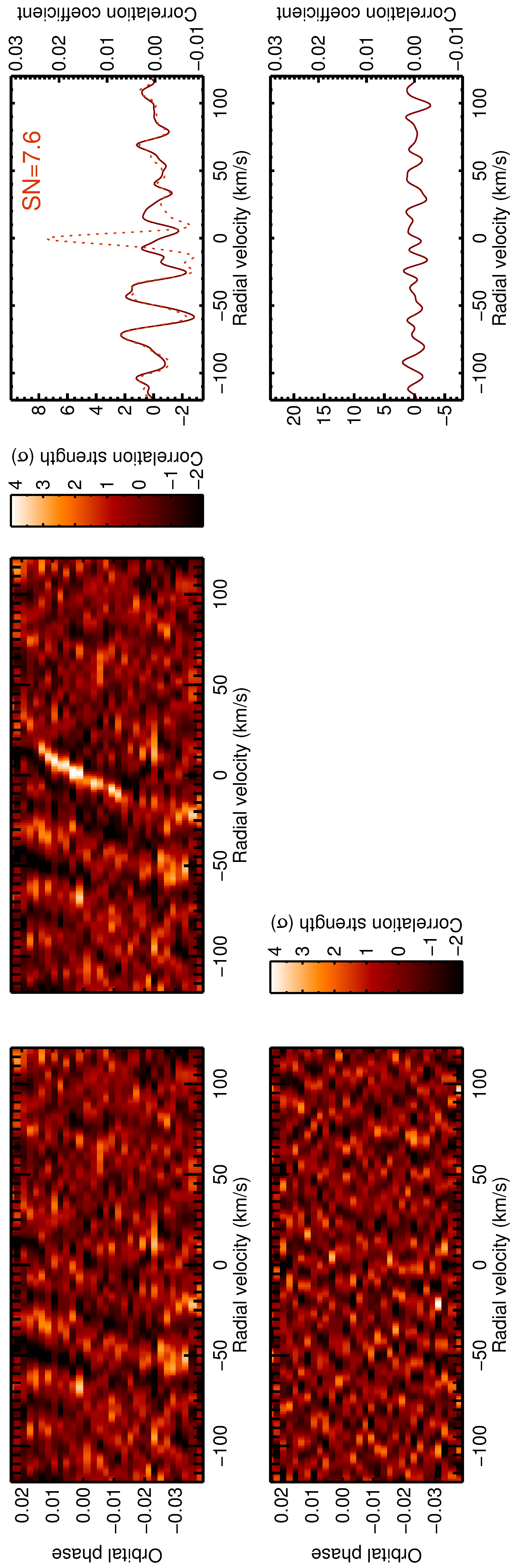}
   \caption{Top panels: Correlation coefficients for cross-correlation with the TiO template ($\textrm{VMR}=10^{-7}$) between $640 \nm$ and $670 \nm$, with and without injecting the template into the data prior to cross-correlation (middle and left panel, dashed and solid lines respectively). Restricting the wavelength range has reduced the signal-to-noise at which this template is retrieved after injection by over $9 \sigma$ (compare to Figure \ref{xcorresult}), but it is not clear to what extent persistent inaccuracies still adversely affect the sensitivity of this restricted template. Bottom row: Correlation coefficients for cross-correlation Barnard's star's optical spectrum. No enhanced correlation is observed. Given that the injection of the stellar absorption spectrum into the planetary transmission spectrum is not physically meaningful, the sensitivity of this non-detection can't straightforwardly be expressed in physical quantities like the limiting VMR.}%
\label{fig:alternative_correlation}
\end{figure*}

As shown above, the cross-correlation procedure is very sensitive to species with rich optical absorption spectra in the HDS residuals. As for TiO, the sensitivity reaches down to VMRs below the ppb level. Although we have also shown that the line-lists which lie at the basis of our TiO (and other's) model spectra display extensive inaccuracies, the template is able to partially retrieve TiO features from Barnard's star's optical spectrum at wavelengths longwards of $\sim 640 \nm$ and is thus to a certain extent useful to probe the atmosphere of HD 209458b for the presence of TiO. We therefore cross-correlate the residuals with the TiO template at a VMR of $10^{-7}$, using only wavelengths between $640\nm - 682\nm$, the result of which is shown in the top row of figure \ref{fig:alternative_correlation}.

Injecting the template into the residuals like before, reveals a theoretical sensitivity of $7.6 \sigma$ (Top right panel of Figure \ref{fig:alternative_correlation}). However, because the line list is only partially accurate at these wavelengths, this is only an upper limit: I.e. if line list was perfect, the presence of TiO in the atmosphere of HD 209458b at a volume-mixing ratio of $10^{-7}$ would have been retrieved with a significance of $7.6 \sigma$. Though since the line-list suffers from uncharacterised inaccuracies, the true limiting VMR is not determined.

Earlier it has been shown (Figure \ref{xcorcompare}) that if the temperature-pressure environment of the TiO template does not match that of the data, the correlation is not severely affected. Therefore, although the temperature of the photosphere of Barnard's star is some $1500\textrm{K}$ higher than the atmosphere of HD 209458b at the terminator region, the TiO absorption present in the spectra of M-dwarfs may also be useful to identify TiO absorption in exoplanet atmosphere at high resolution. As shown before (Figure \ref{Mdwarf_mutual}), the presence of noise in a cross-correlation template does not destroy the cross-correlation at high-resolution, and thus we propose to use the HARPS spectrum of Barnard's star as a template for cross-correlation with the HDS data of HD 209458b.

Injecting the absorption spectrum of Barnard's star into the residuals of the HDS data does not provide a physically meaningful sensitivity. Although our template at $\sim 1500\textrm{K}$ correlates significantly with the stellar atmosphere models, the absorption spectrum of an M-dwarf is not a good model for the absorption spectrum of the upper atmosphere of HD 209458b during transit, because the underlying physical parameters are very different: In addition to the different temperature-pressure environment, the presence of other species in an M-dwarf's photosphere and the different continuum level (HD 209458 is a solar type star) make it difficult to adjust Barnard's star's absorption spectrum to form a template from which parameters such as the limiting VMR can be derived as before. Extracting a physically meaningful TiO absorption spectrum from real-world absorption spectra of M-dwarfs would require more work and is beyond the scope of this paper. For now, we only use Barnard's star's spectrum to qualitatively probe for the presence of TiO absorption in the HDS dataset. The resulting cross-correlation is shown in Figure \ref{fig:alternative_correlation}. Again, no significant correlation is detected.


\section{Conclusion}
We have probed the optical transmission spectrum of HD 209458b for TiO absorption using the time-dependent radial velocity of the planet, which can be spectrally resolved at high resolution. At first glance our null-result seems to be in contrast with the possible detection of TiO by \citet{Desert2008}, and in line with more recent observations of hot-Jupiters and models. Injection of model TiO spectra into the data shows that our cross-correlation retrieval method is sensitive to volume mixing ratios down to $10^{-10}$, and is thus theoretically capable of ruling out the presence of TiO at a stringent level. 

However, this method relies on the accuracy of template spectra of TiO at resolutions in the order of $R \sim 10^5$. Although such templates are available in the literature, we observe that the line databases used to synthesise these spectra are based on model calculations of the TiO molecule which display widespread inaccuracies. We infer that the energy levels of the TiO molecule are not determined well enough to accurately synthesise absorption spectra at these resolutions.
Nonetheless, we have found that the TiO line list used to generate our template spectrum is accurate to some extent at wavelengths between $640 \nm$ and $670 \nm$, and that the spectra of M-dwarfs can also serve as cross-correlation templates. However, using a high-resolution spectrum of Barnard's star and restricting our template to the aforementioned wavelength range, do not result in a significant correlation. It would therefore seem that the presence of a large concentration of TiO in the upper atmosphere of HD 209458b at the terminator region is unlikely. However, due to the uncertainties regarding the physical meaning of these template models, such a conclusion is tentative.

The lack of an accurate TiO line list is thus a critical hindrance in the application of this retrieval technique to search for the presence of TiO in the optical transmission spectrum of HD 209458b. Also, since the TiO molecule is among the best characterized molecules in astrophysics, we suspect that the absorption spectra of less common and more complicated species are likely to be even less accurately characterized. Furthermore, inaccuracies in the calculations of line positions do not only severely hinder the application of high-resolution retrieval techniques like the one employed in this work, but they also pervade widely used stellar atmosphere models. Modern-day stellar atmosphere codes employ highly advanced modelling techniques to poor line databases, which in the case of TiO is over a decade old.

Therefore we conclude that an increasing community of observational and theoretical astronomers would benefit from an immediate and extensive effort in improving the quality of absorption spectra of molecules like TiO, either through more sophisticated theoretical calculations or by means of direct measurements in the laboratory.


\begin{acknowledgements}
      This work is part of the research programmes PEPSci and VICI 639.043.107, which are financed by the Netherlands Organisation for Scientific Research (NWO), and was also supported by the Leiden Observatory Huygens Fellowship. This work was performed in part under contract with the California Institute of Technology (Caltech)/Jet Propulsion Laboratory (JPL) funded by NASA through the Sagan Fellowship Program executed by the NASA Exoplanet Science Institute. Support for this work was provided in part by NASA through Hubble Fellowship grant HST-HF2-51336 awarded by the Space Telescope Science Institute. Also, this research has made use of the Exoplanet Orbit Database and the Exoplanet Data Explorer at exoplanets.org. Finally, we would like to thank Leigh Fetcher, Jaemin Lee and Patrick Irwin for providing the TiO line-list necessary to perform this analysis.
\end{acknowledgements}

\bibliographystyle{aa} 
\bibliography{MMP} 

\begin{thebibliography}{40}
\expandafter\ifx\csname natexlab\endcsname\relax\def\natexlab#1{#1}\fi

\bibitem[{{Asplund} {et~al.}(2006){Asplund}, {Grevesse}, \& {Jacques
  Sauval}}]{Asplund2006}
{Asplund}, M., {Grevesse}, N., \& {Jacques Sauval}, A. 2006, Nuclear Physics A,
  777, 1

\bibitem[{{Bento} {et~al.}(2014){Bento}, {Wheatley}, {Copperwheat}, {Fortney},
  {Dhillon}, {Hickman}, {Littlefair}, {Marsh}, {Parsons}, \&
  {Southworth}}]{Bento2014}
{Bento}, J., {Wheatley}, P.~J., {Copperwheat}, C.~M., {et~al.} 2014, \mnras,
  437, 1511

\bibitem[{{Birkby} {et~al.}(2013){Birkby}, {de Kok}, {Brogi}, {de Mooij},
  {Schwarz}, {Albrecht}, \& {Snellen}}]{Birkby2013}
{Birkby}, J.~L., {de Kok}, R.~J., {Brogi}, M., {et~al.} 2013, \mnras, 436, L35

\bibitem[{{Brogi} {et~al.}(2014){Brogi}, {de Kok}, {Birkby}, {Schwarz}, \&
  {Snellen}}]{Brogi2014}
{Brogi}, M., {de Kok}, R.~J., {Birkby}, J.~L., {Schwarz}, H., \& {Snellen},
  I.~A.~G. 2014, \aap, 565, A124

\bibitem[{{Brogi} {et~al.}(2012){Brogi}, {Snellen}, {de Kok}, {Albrecht},
  {Birkby}, \& {de Mooij}}]{Brogi2012}
{Brogi}, M., {Snellen}, I.~A.~G., {de Kok}, R.~J., {et~al.} 2012, \nat, 486,
  502

\bibitem[{{Brogi} {et~al.}(2013){Brogi}, {Snellen}, {de Kok}, {Albrecht},
  {Birkby}, \& {de Mooij}}]{Brogi2013}
{Brogi}, M., {Snellen}, I.~A.~G., {de Kok}, R.~J., {et~al.} 2013, \apj, 767, 27

\bibitem[{{Burrows} {et~al.}(2007){Burrows}, {Hubeny}, {Budaj}, {Knutson}, \&
  {Charbonneau}}]{Burrows2007}
{Burrows}, A., {Hubeny}, I., {Budaj}, J., {Knutson}, H.~A., \& {Charbonneau},
  D. 2007, \apjl, 668, L171

\bibitem[{{Burrows} \& {Orton}(2009)}]{Burrows2009}
{Burrows}, A. \& {Orton}, G. 2009, ArXiv e-prints

\bibitem[{{Castelli} \& {Kurucz}(2003)}]{Castelli2003}
{Castelli}, F. \& {Kurucz}, R.~L. 2003, in IAU Symposium, Vol. 210, Modelling
  of Stellar Atmospheres, ed. N.~{Piskunov}, W.~W. {Weiss}, \& D.~F. {Gray},
  20P

\bibitem[{{Crossfield} {et~al.}(2012){Crossfield}, {Knutson}, {Fortney},
  {Showman}, {Cowan}, \& {Deming}}]{Crossfield2012}
{Crossfield}, I.~J.~M., {Knutson}, H., {Fortney}, J., {et~al.} 2012, \apj, 752,
  81

\bibitem[{{de Kok} {et~al.}(2013){de Kok}, {Brogi}, {Snellen}, {Birkby},
  {Albrecht}, \& {de Mooij}}]{DeKok2013}
{de Kok}, R.~J., {Brogi}, M., {Snellen}, I.~A.~G., {et~al.} 2013, \aap, 554,
  A82

\bibitem[{{D{\'e}sert} {et~al.}(2008){D{\'e}sert}, {Vidal-Madjar}, {Lecavelier
  Des Etangs}, {Sing}, {Ehrenreich}, {H{\'e}brard}, \& {Ferlet}}]{Desert2008}
{D{\'e}sert}, J.-M., {Vidal-Madjar}, A., {Lecavelier Des Etangs}, A., {et~al.}
  2008, \aap, 492, 585

\bibitem[{{Diamond-Lowe} {et~al.}(2014){Diamond-Lowe}, {Stevenson}, {Bean},
  {Line}, \& {Fortney}}]{Diamond-Lowe2014}
{Diamond-Lowe}, H., {Stevenson}, K.~B., {Bean}, J.~L., {Line}, M.~R., \&
  {Fortney}, J.~J. 2014, ArXiv e-prints

\bibitem[{{Fortney} {et~al.}(2010){Fortney}, {Shabram}, {Showman}, {Lian},
  {Freedman}, {Marley}, \& {Lewis}}]{Fortney2010}
{Fortney}, J.~J., {Shabram}, M., {Showman}, A.~P., {et~al.} 2010, \apj, 709,
  1396

\bibitem[{{Freedman} {et~al.}(2008){Freedman}, {Marley}, \&
  {Lodders}}]{Freedman2008}
{Freedman}, R.~S., {Marley}, M.~S., \& {Lodders}, K. 2008, \apjs, 174, 504

\bibitem[{{Gibson} {et~al.}(2013){Gibson}, {Aigrain}, {Barstow}, {Evans},
  {Fletcher}, \& {Irwin}}]{Gibson2013}
{Gibson}, N.~P., {Aigrain}, S., {Barstow}, J.~K., {et~al.} 2013, \mnras, 436,
  2974

\bibitem[{{Hubeny} {et~al.}(2003){Hubeny}, {Burrows}, \&
  {Sudarsky}}]{Hubeny2003}
{Hubeny}, I., {Burrows}, A., \& {Sudarsky}, D. 2003, \apj, 594, 1011

\bibitem[{{Huitson} {et~al.}(2013){Huitson}, {Sing}, {Pont}, {Fortney},
  {Burrows}, {Wilson}, {Ballester}, {Nikolov}, {Gibson}, {Deming}, {Aigrain},
  {Evans}, {Henry}, {Lecavelier des Etangs}, {Showman}, {Vidal-Madjar}, \&
  {Zahnle}}]{Huitson2013}
{Huitson}, C.~M., {Sing}, D.~K., {Pont}, F., {et~al.} 2013, \mnras, 434, 3252

\bibitem[{{Husser} {et~al.}(2013){Husser}, {Wende-von Berg}, {Dreizler},
  {Homeier}, {Reiners}, {Barman}, \& {Hauschildt}}]{Husser2013}
{Husser}, T.-O., {Wende-von Berg}, S., {Dreizler}, S., {et~al.} 2013, \aap,
  553, A6

\bibitem[{{Knutson} {et~al.}(2008){Knutson}, {Charbonneau}, {Allen}, {Burrows},
  \& {Megeath}}]{Knutson2008}
{Knutson}, H.~A., {Charbonneau}, D., {Allen}, L.~E., {Burrows}, A., \&
  {Megeath}, S.~T. 2008, \apj, 673, 526

\bibitem[{{Knutson} {et~al.}(2007){Knutson}, {Charbonneau}, {Noyes}, {Brown},
  \& {Gilliland}}]{Knutson2007}
{Knutson}, H.~A., {Charbonneau}, D., {Noyes}, R.~W., {Brown}, T.~M., \&
  {Gilliland}, R.~L. 2007, \apj, 655, 564

\bibitem[{{Lockwood} {et~al.}(2014){Lockwood}, {Johnson}, {Bender}, {Carr},
  {Barman}, {Richert}, \& {Blake}}]{Lockwood2014}
{Lockwood}, A.~C., {Johnson}, J.~A., {Bender}, C.~F., {et~al.} 2014, \apjl,
  783, L29

\bibitem[{{Mazeh} {et~al.}(2000){Mazeh}, {Naef}, {Torres}, {Latham}, {Mayor},
  {Beuzit}, {Brown}, {Buchhave}, {Burnet}, {Carney}, {Charbonneau}, {Drukier},
  {Laird}, {Pepe}, {Perrier}, {Queloz}, {Santos}, {Sivan}, {Udry}, \&
  {Zucker}}]{Mazeh2000}
{Mazeh}, T., {Naef}, D., {Torres}, G., {et~al.} 2000, \apjl, 532, L55

\bibitem[{{Narita} {et~al.}(2005){Narita}, {Suto}, {Winn}, {Turner}, {Aoki},
  {Leigh}, {Sato}, {Tamura}, \& {Yamada}}]{Narita2005}
{Narita}, N., {Suto}, Y., {Winn}, J.~N., {et~al.} 2005, \pasj, 57, 471

\bibitem[{{Nidever} {et~al.}(2002){Nidever}, {Marcy}, {Butler}, {Fischer}, \&
  {Vogt}}]{Nidever2002}
{Nidever}, D.~L., {Marcy}, G.~W., {Butler}, R.~P., {Fischer}, D.~A., \& {Vogt},
  S.~S. 2002, \apjs, 141, 503

\bibitem[{{Parmentier} {et~al.}(2013){Parmentier}, {Showman}, \&
  {Lian}}]{Parmentier2013}
{Parmentier}, V., {Showman}, A.~P., \& {Lian}, Y. 2013, \aap, 558, A91

\bibitem[{{Plez}(1998)}]{Plez1998}
{Plez}, B. 1998, \aap, 337, 495

\bibitem[{{Portmann} \& {Solomon}(2007)}]{Portmann2007}
{Portmann}, R.~W. \& {Solomon}, S. 2007, \grl, 34, 2813

\bibitem[{{Rodler} {et~al.}(2013){Rodler}, {K{\"u}rster}, \&
  {Barnes}}]{Rodler2013}
{Rodler}, F., {K{\"u}rster}, M., \& {Barnes}, J.~R. 2013, \mnras, 432, 1980

\bibitem[{{Rodler} {et~al.}(2012){Rodler}, {Lopez-Morales}, \&
  {Ribas}}]{Rodler2012}
{Rodler}, F., {Lopez-Morales}, M., \& {Ribas}, I. 2012, \apjl, 753, L25

\bibitem[{{Rowe} {et~al.}(2006){Rowe}, {Matthews}, {Seager}, {Kuschnig},
  {Guenther}, {Moffat}, {Rucinski}, {Sasselov}, {Walker}, \&
  {Weiss}}]{Rowe2006}
{Rowe}, J.~F., {Matthews}, J.~M., {Seager}, S., {et~al.} 2006, \apj, 646, 1241

\bibitem[{{Schwenke}(1998)}]{Schwenke1998}
{Schwenke}, D.~W. 1998, Faraday Discussions, 109, 321

\bibitem[{{Sing} {et~al.}(2013){Sing}, {Lecavelier des Etangs}, {Fortney},
  {Burrows}, {Pont}, {Wakeford}, {Ballester}, {Nikolov}, {Henry}, {Aigrain},
  {Deming}, {Evans}, {Gibson}, {Huitson}, {Knutson}, {Showman}, {Vidal-Madjar},
  {Wilson}, {Williamson}, \& {Zahnle}}]{Sing2013}
{Sing}, D.~K., {Lecavelier des Etangs}, A., {Fortney}, J.~J., {et~al.} 2013,
  \mnras, 436, 2956

\bibitem[{{Snellen} {et~al.}(2008){Snellen}, {Albrecht}, {de Mooij}, \& {Le
  Poole}}]{Snellen2008}
{Snellen}, I.~A.~G., {Albrecht}, S., {de Mooij}, E.~J.~W., \& {Le Poole}, R.~S.
  2008, \aap, 487, 357

\bibitem[{{Snellen} {et~al.}(2010){Snellen}, {de Kok}, {de Mooij}, \&
  {Albrecht}}]{Snellen2010}
{Snellen}, I.~A.~G., {de Kok}, R.~J., {de Mooij}, E.~J.~W., \& {Albrecht}, S.
  2010, \nat, 465, 1049

\bibitem[{{Spiegel} {et~al.}(2009){Spiegel}, {Silverio}, \&
  {Burrows}}]{Spiegel2009}
{Spiegel}, D.~S., {Silverio}, K., \& {Burrows}, A. 2009, in AAS/Division for
  Planetary Sciences Meeting Abstracts, Vol.~41, AAS/Division for Planetary
  Sciences Meeting Abstracts 41, 68.03

\bibitem[{{Stevenson} {et~al.}(2014){Stevenson}, {Bean}, {Seifahrt},
  {D{\'e}sert}, {Madhusudhan}, {Bergmann}, {Kreidberg}, \&
  {Homeier}}]{Stevenson2014}
{Stevenson}, K.~B., {Bean}, J.~L., {Seifahrt}, A., {et~al.} 2014, \aj, 147, 161

\bibitem[{{Torres} {et~al.}(2008){Torres}, {Winn}, \& {Holman}}]{Torres2008}
{Torres}, G., {Winn}, J.~N., \& {Holman}, M.~J. 2008, \apj, 677, 1324

\bibitem[{{Wright} {et~al.}(2011){Wright}, {Fakhouri}, {Marcy}, {Han}, {Feng},
  {Johnson}, {Howard}, {Fischer}, {Valenti}, {Anderson}, \&
  {Piskunov}}]{Wright2011}
{Wright}, J.~T., {Fakhouri}, O., {Marcy}, G.~W., {et~al.} 2011, \pasp, 123, 412

\bibitem[{{Zellem} {et~al.}(2014){Zellem}, {Lewis}, {Knutson}, {Griffith},
  {Showman}, {Fortney}, {Cowan}, {Agol}, {Burrows}, {Charbonneau}, {Deming},
  {Laughlin}, \& {Langton}}]{Zellem2014}
{Zellem}, R.~T., {Lewis}, N.~K., {Knutson}, H.~A., {et~al.} 2014, \apj, 790, 53

\end{thebibliography}

\end{document}